\documentclass[conference]{IEEEtran}
\IEEEoverridecommandlockouts
\usepackage{cite}
\usepackage{amsmath,amssymb,amsfonts}
\usepackage{algorithmicx}
\usepackage{graphicx}
\usepackage{textcomp}
\usepackage{xcolor}
\definecolor{mygray}{gray}{.9}
\usepackage{graphicx}
\usepackage{float}
\usepackage{subfigure}
\usepackage{textcomp}
\usepackage{tabularx}
\usepackage{algorithm}
\usepackage{algpseudocode}
\usepackage{amsmath}
\usepackage{multirow}
\usepackage{verbatim}
\usepackage{colortbl}
\usepackage{enumerate}
\usepackage{hyperref}
\hypersetup{hypertex=true,
	colorlinks=true,
	linkcolor=black,
	anchorcolor=black,
	citecolor=black}
\usepackage{threeparttable}
\usepackage[misc]{ifsym}
\def\BibTeX{{\rm B\kern-.05em{\sc i\kern-.025em b}\kern-.08em
		T\kern-.1667em\lower.7ex\hbox{E}\kern-.125emX}}
\begin{document}
	
	\title{MAAC: Novel Alert Correlation Method To Detect Multi-step Attack\\
		\thanks{This work was supported by the National Natural Science Foundation of China (Grant No. 61802394 and 61902396) and the Youth Innovation Promotion Association. This work is also supported by the Program of Key Laboratory of Network Assessment Technology, the Chinese Academy of Sciences and Program of Beijing Key Laboratory of Network Security and Protection Technology.}
	}

	\author{%
		\IEEEauthorblockN{Xiaoyu Wang\IEEEauthorrefmark{1}\IEEEauthorrefmark{2}, Xiaorui Gong\IEEEauthorrefmark{1}\IEEEauthorrefmark{2}\Letter, Lei Yu\IEEEauthorrefmark{1}\IEEEauthorrefmark{2}, Jian Liu\IEEEauthorrefmark{1}\IEEEauthorrefmark{2}
			\IEEEauthorblockA{\IEEEauthorrefmark{1}School of Cyber Security, University of Chinese Academy of Sciences, Beijing, China}
			\IEEEauthorblockA{\IEEEauthorrefmark{2}Institute of Information Engineering, Chinese Academy of Sciences, Beijing, China}			
			Email: \{wangxiaoyu, gongxiaorui, yulei, liujian6\}@iie.ac.cn}\\
	}
	

	\maketitle
	
	\begin{abstract}
		With the continuous improvement of attack methods, there are more and more distributed, complex, targeted attacks in which the attackers use combined attack methods to achieve the purpose. Advanced cyber attacks include multiple stages to achieve the ultimate goal. Traditional intrusion detection systems such as endpoint security management tools, firewalls, and other monitoring tools generate a large number of alerts during the attack. These alerts include attack clues, as well as many false positives unrelated to attacks. Security analysts need to analyze a large number of alerts and find useful clues from them and reconstruct attack scenarios. However, most traditional security monitoring tools cannot correlate alerts from different sources, so many multi-step attacks are still completely unnoticed, requiring manual analysis by security analysts like finding a needle in a haystack. We propose MAAC, a multi-step attack alert correlation system, which reduces repeated alerts and combines multi-step attack paths based on alert semantics and attack stages. The evaluation results of the real-world datasets show that MAAC can effectively reduce the alerts by 90\% and find attack paths from a large number of alerts.
	\end{abstract}
	
	
	\section{Introduction}
	With the continuous development of cyberspace, attacks have become more complex. Generally speaking, "attack" refers to a single-step attack, also known as an atomic attack, that is, an inseparable atomic malicious behavior, e.g., SQL injection, Cross-Site Scripting (XSS). However, with the complexity of the network structure and the enhancement of defense capabilities of the defender, attack targets (such as database servers and programmable logic controllers, etc.) are usually located in the internal network and are protected by multiple security products. In this case, the attacker often cannot achieve the attack purpose through a single simple step. Combining multiple attack steps is needed to achieve the attacker's purpose. This kind of attack is called a multi-step attack, also called a multi-stage attack or Multistage attack. The characteristic of a multi-step attack is that the attack process includes at least two atomic attacks, and there is a logical relationship between the two atomic attacks. Compared with traditional single-step attacks, multi-step attacks are more complicated, have a longer period, and are more concealed. Discovering the behavior of attackers through various information and understanding the security situation of the entire network is the focus of multi-step attack detection.
	
	To ensure the security of the network, a great deal of security protection equipment and software are deployed in the network, for example, border firewall equipment, intrusion detection system (IDS), and anti-virus software. According to the survey \cite{Report1}, among 750 network security and IT professionals, 78\% rely on more than 50 security products to solve security problems, and 37\% use more than 100 security tools. Most of these security products use traditional feature matching or statistical analysis methods to detect known attacks, which will generate a large number of false positives. The report \cite{Report2} points out that more than 37\% of companies generate more than 10,000 alerts each month, 52\% of alerts are false positive, and 36\% are redundant across multiple threat detection platforms.
	
	During the multi-step attack process, the IDSs which are used to protect the entire network endpoints generate a large number of alerts because of the long time attack process and multiple attack steps. Besides, an attack action may be performed frequently or trigger different sensors \cite{haas2018gac:}, which also increases the number of alerts. These alerts are scattered, indicating that a certain step of the attack has been detected. However, alerts cannot describe the overall attack scenario. It takes security analysts a lot of time and effort to find effective alerts from a huge number of alerts, and then dig out the logical relationship between alerts, and reconstruct attack scenarios from these alerts. Therefore, it is necessary to automatically integrate and correlate alerts, discover the attack process, and reconstruct the attack scenario.
	
	However, the existing alert correlation algorithm \cite{haas2018gac:,julisch2001mining,Klaus2003Clustering,2006Alert} correlates alerts based on characteristics, which ignores the intrinsic relationship of alerts and cannot perform semantic correlation. GAC \cite{haas2018gac:} clusters the alerts and classifies alerts into one of four types. Then the clusters are connected according to the communication between the hosts to compose a complete multi-step attack. However, GAC can only identify pre-defined attack scenarios, there are still many attack types such as vulnerability exploits that cannot be identified. GAC ignores the alert information which is already given by IDS, and only uses IP, port, and protocol as the indicators. Ning et al. propose an alert correlation algorithm \cite{ning2002constructing,ning2002analyzing} which uses pre-defined alert correlation rules, but this method can only identify known attacks.
	Traditional Security Information and Event Management (SIEM) correlates alerts based on existing rules. The rules are manually formulated and can only detect existing attacks.
	
	\textbf{Problem Statement.} \textit{The main problem solved in this paper is to understand the semantics of alerts, aggregate and correlate alerts through semantics, and get the order of suspicious hosts and attack alerts paths.} There are three main aspects to this problem, and they are as follows:
	
	\textbf{Alert semantics.} The algorithm or program should understand the undermeaning of the alert, and obtain the attack stage of the host through alert semantics analysis. Understanding the alert semantics can help alert aggregation and correlation. 
	
	\textbf{Attack correlation.} The challenge is to automatically correlate alerts through logical semantics instead of correlation through feature matching or manual rules.  
	
	\textbf{Attack path discovery.} The goal is to build an attack alert graph and provide possible attack scenarios. The attack scenario graph or path helps security analysts accurately discover multi-step attacks. Besides, to better explain the consequences of the attack, it is necessary to rank the suspicious hosts. 
	
	\textbf{Our Approach.} In this paper, we use the information field of the alert to correlate the alerts and discover the multi-step attack. We first vectorize the information field of the alert and aggregate similar alerts into one super alert to reduce the duplication of alerts. Meanwhile, the alert information is used to label the attack stage to which the alert belongs. Then the alerts belonging to different attack stages are connected to form a complete attack process. Our method has been evaluated on both the alert classification test and three real-world datasets. The results show that our method can reduce a large number of repeated alerts by more than 90\%, compose alerts into an alert correlation graph, and effectively extract the most suspicious attack path from the graph. In summary, our contributions are as follows: 
	\begin{itemize}
		\item We introduce a novel alert correlation algorithm that could understand alert semantics. 
		\item We propose MAAC$\footnote{MAAC for \textbf{M}ulti-step \textbf{A}ttack detection by \textbf{A}lert \textbf{C}orrelation.}$, a prototype system for alert aggregation, alert correlation, and alert graph generation. MAAC uses semantics to classify alerts, find attack paths, and rank suspicious hosts.
		\item We conduct the evaluation of MAAC in similarity comparison results and the attack scenario path discovery results. Compared to a state-of-the-art alert correlation algorithm,  MAAC can aggregate alerts better, and MAAC does not need correlation rules in advance.
	\end{itemize}
	
	The remainder of this paper is structured as follows: Section \ref{sec:realtedwork} summarizes related work. Section \ref{sec:alertcorr} introduces the threat model, alert reduction, and alert correlation method. In section \ref{sec:evaluation}, we verify the semantic understanding ability and alerts correlation result on three datasets. Section \ref{sec:conclusion} concludes our work.
	
	\section{Related WORK}
	\label{sec:realtedwork}
	
	\subsection{Alert Correlation}
	
	Alert-based multi-step attack detection can be divided into five parts: attribute similarity-based, interconnection-based, statistics-based, Structure-based, and provenance-based.
	
	\textbf{Attribute similarity-based.} Similarity clustering \cite{haas2018gac:,julisch2001mining,Klaus2003Clustering,2006Alert} is to cluster similar alerts through attribute similarity detection. This method is based on the principle that alerts shares with similar attributes usually belong to the same attack. Similarly, Haas et al. \cite{EfficientAttackCorrelation} clusters alert and convert them into motif signatures which are the abstract description of the attack.
	
	\textbf{Interconnection-based.} The inner connection is the logical connection relationship of alerts. In approaches \cite{ning2002constructing,ning2002analyzing}, alerts are correlated based on pre-defined attack prerequisites and consequences. This method correlates the alerts that respectively satisfy the prerequisites and consequences conditions, and can find the logical connection between the alerts. MLAPT \cite{MLAPT} is a machine learning-based system that implements eight detection modules to discover different APT steps and then correlates the alerts of the detection modules followed by the prediction module. 
	
	\textbf{Statistics-based.} These approaches discover the logical relationship between alerts through statistics and inference. Qin et al. \cite{qin2003statistical} uses the Granger Causality Test, a time series statistical causality analysis, to alerts correlation. Statistics-based alert correlation does not require prior knowledge about the attack information. \cite{qin2004discovering} applies Bayesian network for alert correlation, which outputs the alert correlation possibility. 
	
	\textbf{Structure-based.} The attackers perform attack actions according to the target network structure and the value of the assets in the network, so the attacker's actions can be planned or predicted by abstracting the network structure. Approaches \cite{Chien2012,noel2004correlating,Wang2006,Zhang2008} map alert to attack graph or attack tree nodes to correlate alert events. Structure-based methods require prior knowledge of network topology and asset information.
	
	\textbf{Provenance-based.} Data provenance graph is a directed graph called the provenance graph, which describes data flow between system subjects (e.g., processes) and objects (e.g., files and sockets). NoDoze \cite{hassan2019nodoze} combines alert correlation with data provenance to solve the problem of alert fatigue. NoDoze generates alert dependency graphs providing security analysts with contextual background information. Data provenance records the system runtime status and reveals the relationships between entities. Security analysts can analyze the source and impact of attacks through the provenance graph. However, the provenance-based approaches have the dependence explosion problem.

	\subsection{Natural Language Processing}
	
	Natural Language Processing (NLP) technology understands and analyzes natural language. NLP can be used in speech recognition, question-answering systems, machine translation, etc. With the maturity of this technology, its application in the field of cybersecurity has gradually increased. Binary code similarity comparison uses NLP technology to represent binary code as a vector for similarity comparison. The word2vec model \cite{2013Efficient} can map each word to a vector to represent the relationship between words. Zuo et al. \cite{Neural_Machine_Translation_Binary_Code_2019} regard instructions as words and basic blocks as sentences, and use word2vec to embedding the instruction. The advantage of this method is that it can solve the cross-architecture code containment problems. This research shows that using the methods, ideas, and techniques in NLP to perform the binary analysis is promising.

	\section{Alert Correlation}
	\label{sec:alertcorr}
	
	In this section, we present the approach of our novel alert reduction and correlation algorithm. First, we propose a threat model as an example of a multi-step attack, then give the overview of the algorithm, and finally introduce each module in detail.
	
	\subsection{Threat Model}
	
\begin{figure}[htbp]
		\centering
		\includegraphics[width=0.5\textwidth,trim=0 20 0 10,clip]{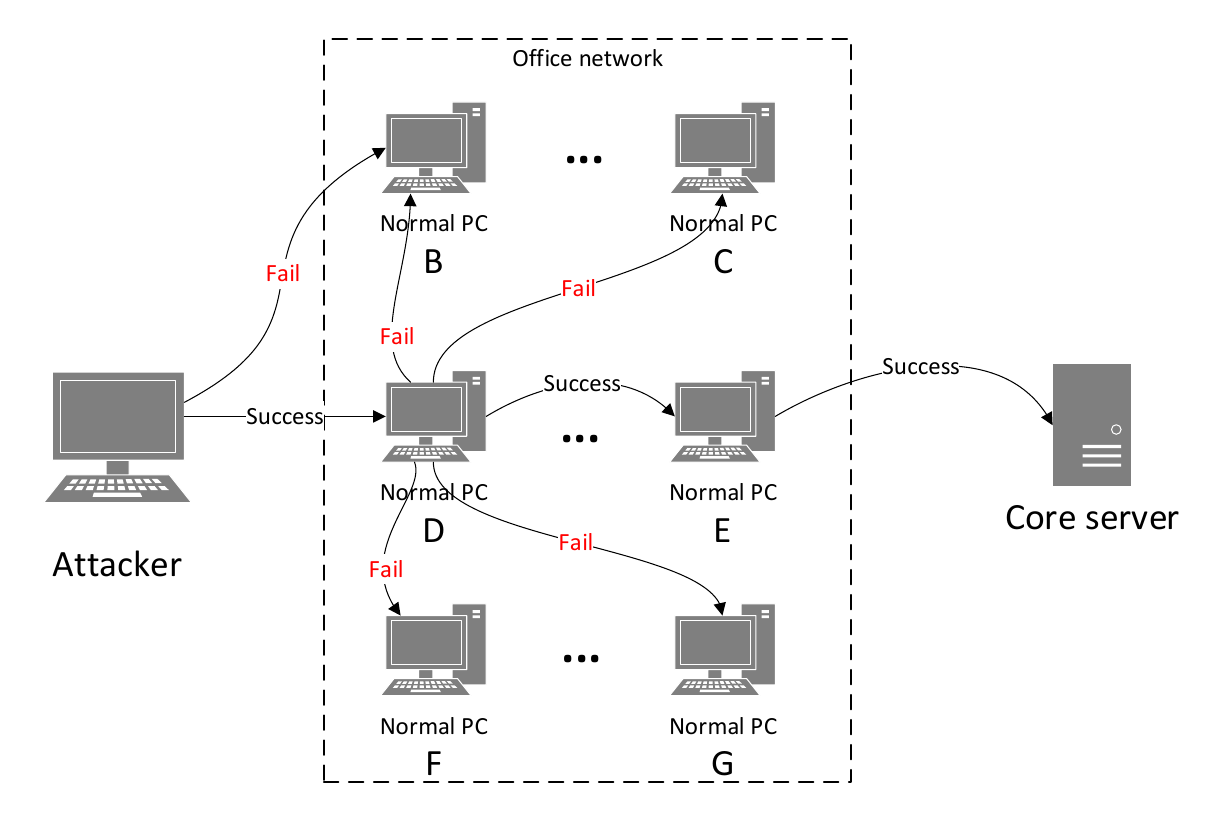}
		\caption{Threat Model}
		\label{fig:threatmodel}
	\end{figure}

	The attacker first attacks two hosts in the office network and exploits the vulnerability such as \textit{cve-2017-11882} \cite{cve-2017-11882} through phishing emails or other methods -- but only one succeeds -- then installs a backdoor and moves laterally in the intranet. This controlled host scans the intranet and discovers vulnerabilities such as \textit{MS17-010} (\textit{Eternalblue}) \cite{ms17010} to exploit the other hosts in the intranet, and only one succeeded to obtain the root privilege of the host. This host is connected to the core server. After that, this host attacks the core server to obtain secret information. The attack process is shown in Figure \ref{fig:threatmodel}. To express intuitively, network devices are omitted, such as routers, switches, and firewalls.
	
	During the attack phase, the boundary and internal network intrusion detection system (NIDS) respectively generate plenty of alerts. Besides, there are a large number of alerts about malicious files and backdoors activities generated by the endpoint security software installed on the hosts. The alerts are messy and there are duplicates. The dispersion and duplication of alerts obstruct security analysts unable to find out real information related to the attack, increasing the difficulty of attack detection. Therefore, it is necessary to integrate repeated alerts, remove irrelevant alerts, and generate attack alert paths.
	
	
	
	\subsection{Overview}
\begin{figure*}[htbp] 
		\centering 
		\includegraphics[width=1\textwidth,trim=0 20 0 10,clip]{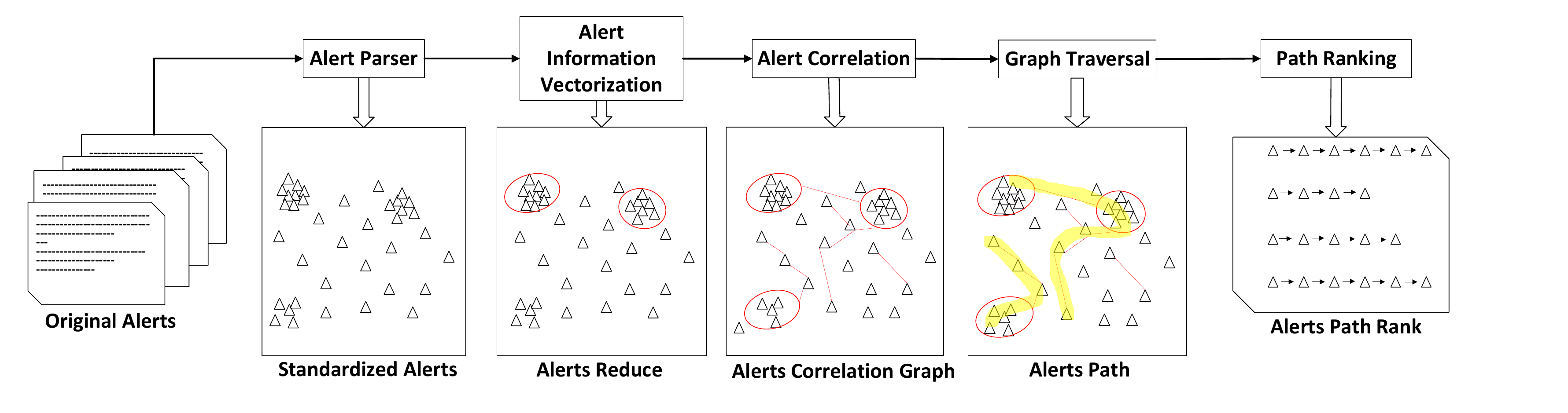} 
		\caption{System Design of MAAC} 
		\label{Fig.systemdesign} 
	\end{figure*}

	In this section, we propose MAAC, a novel method to reduce and correlate the alerts reported by IDS and EDR. It analyzes the semantic representation of alerts, generates a concise alert graph, and traverses to get attack paths. Figure \ref{Fig.systemdesign} describes the key phases of MAAC. Note that our approach focuses on alert processing, and how the alert is generated is beyond the scope of our research.

	First, MAAC preprocesses the alerts and formats the alerts generated by different sensors into a unified pattern. And then it combinates the duplicate alert to reduce the number of alerts. It measures the similarity between alerts and classifies their attack stage according to the alert semantics. Then MAAC generates an alert correlation graph based on the attack stage. Finally, MAAC sorts the nodes in the graph and traverses them to obtain suspicious attack paths and hosts. The input of MAAC is the alerts from multiple sources, while the output of MAAC is the attack paths and hosts ranking. 
	
	\subsection{Alert Reduction}
	
	The novel alert reduction method uses the alert description field, which is the most useful attribute in an alert. The description field includes the reason for triggering this alert. For example, the description field of the Eternalblue attack alert is \textit{Possible ETERNALBLUE MS17-010 Echo Response}. This sentence indicates that the alert belongs to the \textit{echo} feature detection of the Eternalblue attack. There are a large number of sensors and detection tools in the network which generate plenty of alerts. Thus many of these alerts are duplicates. For example, multiple sensors will detect the same exploit and generate alerts with different expressions but similar semantics. Table \ref{alertfields} lists the alert field generated by the general sensor. Timestamp, msg, dst, sensor attributes are in both network and host sensor. Table \ref{threatmodelalert} is the alerts in the threat model. The alert \textit{msg} (also called the feature code) provides a valuable description of the alert, which is the clue that distinguishes the alert from each other. MAAC reduces alerts based on the similarity of alert descriptions. Alerts with a similar \textit{msg} field are more likely to belong to the same attack stage. For example, the vulnerability detection tool \textit{smbtouch} is detected by two NIDS: Snort and Suricata, which are the alert \textit{4} and \textit{9} in Table \ref{threatmodelalert},  respectively. The \textit{msg} of two alerts is different but the undermeaning of these two is the same.

	\begin{table}[htbp]
		\caption{Alert Fields.}
		\centering
		\begin{tabular}{l|l|l}
			\hline
			\rowcolor{mygray}
			fieldname          & description              &alert source  \\ \hline
			timestamp          & Trigger time             &N \& H\\ \hline
			msg                & Alert detail description &N \& H \\ \hline
			src                & Source ip                &N \\ \hline
			dst                & Destination ip           &N \& H \\ \hline    
			proto              & Protocol                 &N \\ \hline 
			dgmlen             & Datagram length          &N \\ \hline 
			srcport            & Source port              &N \\ \hline 
			dstport            & Destination port         &N \\ \hline
			sensor             & Related sensor           &N \& H\\ \hline 
		\end{tabular}
		\label{alertfields}
		\begin{tablenotes} 
			\item {In the alert source column, N=Network sensor, H=Host sensor.} 
		\end{tablenotes} 
	\end{table}
	
	\begin{table*}[htbp]
		\centering
		\begin{threeparttable} 
			\caption{Alerts in Threat Model}
			\begin{tabular}{l|l|l|l|l|l}
				\hline
				\rowcolor{mygray}
				id & timestamp & msg                                                        & src      & dst         & sensor              \\ \hline
				1-2  & time1     & e-mail attachment  maldoc   CVE-2017-11882                 & Attacker & BD           & zeek                \\ \hline
				3  & time2     & malware doc file execute                                   & N/A     & D           & Anti-virus software \\ \hline
				4-8  & time3     & NSA smbtouch scan SMB vulnerable host                      & D        & BCEFG           & Snort               \\ \hline
				9-13  & time3     & fuzzbunch toolset smbtouch test SMB vulnerability          & D        & BCEFG           & Suricata            \\ \hline
				14 & time4     & ETERNALBLUE MS17-010 Echo Request (set)                    & D        & E           & Snort               \\ \hline
				15 & time5     & ETERNALBLUE MS17-010 Echo Response                         & D        & E           & Snort               \\ \hline
				16 & time6     & Possible DOUBLEPULSAR Beacon Response                      & D        & E           & Snort               \\ \hline
				17 & time7     & Apache Struts OGNL Expression Injection (CVE-2017-5638) M2 & E        & Core server & Suricata            \\ \hline
				18 & time8     & TELNET Bad Login                                           & C        & Core server & Snort               \\ \hline
			\end{tabular}
			\label{threatmodelalert}
			\begin{tablenotes} 
				\item \textbf{Due to layout reasons, only meaningful alerts are listed.} 
			\end{tablenotes} 
		\end{threeparttable} 
	\end{table*}
	
	The Paragraph Vector (Doc2vec) \cite{doc2vec}  method is an unsupervised algorithm that can learn from the variable-length text (such as sentences, paragraphs, and documents) to obtain fixed-length feature representations. Doc2vec uses paragraph semantic analysis and considers the order of words, while word2vec only analyzes the meaning of words. We use Doc2vec to get the semantic representation of the alert description. In the training phase, each alert information that all IDS can generate is used as a paragraph in the Doc2vec model. MAAC calculates the cosine distance of the generated vector, and those that exceed the threshold are regarded as different representations of the same security event. 
	
	Alerts of a host can be combined into the same super alert if the following two conditions can be met: 1) consecutive but different alerts of the host in a specified time interval, 2) the information fields of these alerts are similar. When combining the repeated alerts into one super alert, we record the number of repeated alerts as a recommended weight. Alerts generated based on network traffic have the direction from the source host to the target host, so the alert reduction is divided into two steps: reduction in the same direction and reduction between different directions. 
	
	The first layer is the reduction in the same direction. If a host is attacked by the same attacker within a specified time interval and generates consecutive but different alerts, and the information fields of these alerts are similar, then these alerts can be combined into the same super alert. The second layer is the reduction between different directions. Two directions of alerts can be detected in two-way network traffic: attacker exploits to victim or victim replies to the attacker. MAAC aggregates two-way alerts by analyzing the similarity of alert semantics and determines the final direction of the super alert based on the number of similar alerts.

	The alert reduction procedure corresponds to Algorithm \ref{alertreduction}. The function \textit{CHRONOLOGICORDER()} sorts alert set by its timestamp attribute. The msg field of each alert is then vectorized by the \textit{DOC2VEC()} function. The reduction algorithm first aggregates the alerts of the same hosts and direction generated by \textit{GETSAMEDIRECTION()}. The \textit{CALCULATESIMILARITY()} function calculates the similarity of alert pairs. Function \textit{MERGE()} calculates vector distance and aggregates the continuous and similar alerts, and records the number of duplicates. In the second layer of aggregation, function \textit{GETSAMEHOST()} collects alerts between two hosts. Then function \textit{CALCULATESIMILARITY()} calculates the alerts similarity in the same way as the first layer. Finally, \textit{MERGEDIRECTION()} determines the direction of the super alerts and combined alerts.

	
	\begin{algorithm}[htbp]  
		\caption{Alert Reduction Algorithm}  
		\label{alertreduction}
		\begin{algorithmic}[1]
			\Require{Alert Set Generated by Monitor $Set_{allalert}$; Trained Doc2Vec Model $Doc2Vec$;}
			\Ensure{Reduced Alert Set $Set_{reduced}$;}
			\State $Set_{ordered}$ = CHRONOLOGICORDER($Set_{allalert}$);
			\State $Vec_{alert}$ = DOC2VEC($Set_{ordered}$);
			\State $Set_{samedirection}$ = GETSAMEDIRECTION($Set_{ordered}$);
			\For{$Alert_{pairs}$ in $Set_{samedirection}$}
			\State $sim$ = CALCULATESIMILARITY($Vec_{Alert_{pairs}}$);
			\State $Set_{firstreduced}$ += MERGE($Alert_{pairs}$ , $sim$);
			\EndFor
			\State $Set_{twowayalert}$ = GETSAMEHOST($Set_{firstreduced}$);
			\For{$Alert_{pairs}$ in $Set_{twowayalert}$}
			\State $sim$ = CALCULATESIMILARITY($Vec_{Alert_{pairs}}$);
			\State $Set_{reduced}$ += MERGEDIRECTION($Alert_{pairs}$,$sim$);
			\EndFor
		\end{algorithmic}
	\end{algorithm}

	\subsection{Alert Graph Generation}
	
	After integrating repeated alerts, MAAC correlates the alerts and generates an alert graph. The alert graph generation has two stages, first is the correlation of the alerts generated on the same host, and then is the correlation of the alerts between the hosts. 
	
	\textbf{Alerts of the same host}. For the alerts from the same source host, MAAC connects these alerts in chronological order. For example, the attacker will first use SQL injection and then XSS attacks to test whether the target host has vulnerabilities. These attacks should be recorded in the attack scenario and appear in the final attack path. Attacks from the same source host belong to the same attacker and can present the entire attack stage. 
	
	\textbf{Alert across hosts}. After correlating the alerts of the same host, MAAC performs cross-host correlation according to the stage of the alert.
	According to the attack steps, MAAC divides alerts into four attack stages: \textit{scan}, \textit{exploit}, \textit{get-access-privilege}, and \textit{post-attack}. The \textit{scan} is the first step in an attack. The goal is to probe and collect information in various ways to prepare for subsequent attacks. In addition to detection in the traditional method, it also includes collecting information through social engineering. \textit{Exploit} is the specific attack process, including attack through vulnerabilities or running malicious documents on the victim host through users. When the attack is successful, the attacker needs to obtain the permission of the victim host to control it and perform subsequent attacks. Therefore, one of the indicators for judging whether the attack is successful is whether there is permission acquisition. This stage is \textit{get-access-privilege}. After successfully attacking a host, the attacker will carry out follow-up actions such as information theft, which is a \textit{post-attack}. The actions of the attacker at each stage are shown in the following table \ref{alertstage}.
	
	\begin{table}[htbp]
		\caption{Alert stages}
		\begin{tabular}{l|l|p{5cm}}
			\hline
			\rowcolor{mygray} 
			No. &Alert Stage          & Alert type              \\ \hline  
			1         &Scan                 & IP address scan, port scan, version scan, Vulnerability scan, social engineering   \\ \hline  
			2         &Exploit              & Malicious file in network traffic and host, command injection, vulnerability attack          \\ \hline  
			3         &Get-Access-Privilege & SSH login, RDP login, shell connect    \\ \hline  
			4         &Post-Attack          & Data transfer, command\&control, backdoor communication \\ \hline  
		\end{tabular}
		\label{alertstage}
	\end{table}
	
	Assuming that there is an alert $A_1$ of type "Get-Access-Privilege" on the host $H_1$, MAAC records the time when the alert is triggered, which is recorded as the compromised time. For the alert sequence from host $H_1$ to host $H_2$, connect alert $A_1$ with the first alert after the compromised time. The ultimate goal of an attack is to gain control of the host, so there is a process of penetration attempts and root connection with the attacker (if successful). If a host is successfully attacked, all subsequent activities from this host against other hosts can be considered attack activities. If host $H_1$ has multiple alerts to host $H_2$ after compromised time, only the alert with the closest time will be connected. For example, as described in the previous threat model, an attacker will obtain the root authority of the host $D$ after trying various attack methods, and then use this host to attack other hosts. In this case, the first victim host $D$ will generate a "Get-Access-Privilege" type alert. After the alert is issued, subsequent alerts generated from $D$ to $B$ are part of the entire attack process of attacking $B$ after attacking $D$. The algorithm of graph generation is shown in Algorithm \ref{alertgraphgene}. The function \textit{CLASSIFYSTAGE()} assigns stage for each alerts through alert semantics. The same host alert connection is in \textit{INNERHOSTCONNECTION()}. \textit{GETSTAGE3ALERT()} gets the alerts belonging to the third stage, that is, \textit{Get-Access-Privilege} type.
	
	\begin{algorithm}[htbp]  
		\caption{Alert Graph Generation Algorithm}  
		\label{alertgraphgene}
		\begin{algorithmic}[1]
			\Require{Reduced Alert Set $Set_{reduced}$}
			\Ensure{Alert Graph $Graph_{alert}$}
			\State $Set_{classified}$ = CLASSIFYSTAGE($Set_{reduced}$);
			\State $Set_{host}$ = $Set_{classified}$ with Same Destination Host;
			\State $OrderSet_{host}$ = CHRONOLOGICORDER($Set_{host}$);
			\For {$OrderSet_{host_i}$ in $OrderSet_{host}$}
			\State INNERHOSTCONNECTION($OrderSet_{host_i}$);
			\State $Alerti_{stage3}$ = GETSTAGE3ALERT($OrderSet_{host_i}$)
			\For{$Alertj$ in $OrderSet_{host_j}$}
			\If {$Alertj_{time}$ \textgreater  $Alerti_{stage3_{time}}$ and $Alertj_{sip}$ == $host_i$}
			\State ADDEDGE($Alerti_{stage3}$ , $Alertj$);
			\State Break;
			\EndIf
			\EndFor
			\EndFor
		\end{algorithmic}
	\end{algorithm}

	\subsection{Alert Path Score Calculation}
	
	After generating the alert graph, MAAC finds all paths with three or more nodes by traversing the graph. The path with too few nodes(less or equal to two) is just a one-step attack, which does not need to be correlated and exceeds the scope of our research. There may be multiple paths in the alert graph, we propose a series of equations calculating the probability of each path to get the most likely attack path.

	\textbf{Host Suspiciousness}. Determining the accuracy of the path requires the reliability of the alert. The reliability of the alert is related to the type of alert, suspicious state of the source host, and destination host. Therefore, the suspiciousness of the host needs to be calculated first. The PageRank algorithm \cite{page1999pagerank} uses the number and quality of hyperlinks between web pages as the main factors to analyze the importance of web pages. MAAC regards each host as a web page and uses alert as the directed edge of the link between the hosts. The basic assumption of PageRank is: the important pages are often quoted more by other pages, that is, there will be more hyperlinks to this important page from other pages. For the host, the host under more attacks has more alerts connected to it, that is, the suspiciousness score is higher.
	
	The suspiciousness of the host is also related to the type of alert. MAAC uses the number of alert types on the host as one of the indicators to evaluate the suspiciousness of the host. When there are multiple types of alerts on a host, the host is more likely to be attacked or even successfully attacked. 
	
	The host suspiciousness is calculated in \eqref{suspicioushost}. $PageRank(h)$ is the PageRank result of the host, and $|AlertType_{h}|$ is the number of alert types on the host.
	
	\begin{equation}
	\label{suspicioushost}
	Suspicious_{h} = PageRank(h) + |AlertType_{h}|
	\end{equation}
	
	\textbf{Alert Reliability}. After evaluating the suspiciousness of the host, we can evaluate the reliability of the alert. In \eqref{reliabilityalert}, $Suspicious_{shost}$ and $Suspicious_{dhost}$ represent the suspiciousness of the source host and destination host of the alert, respectively. It is notable that when the alert is generated by host-based security tools such as EDR and HIDS, there is no source host. $AlertType$ is the original score of the alert. According to the severity of different types, alerts have original values. The initial value of \textit{Scan} is 1, \textit{Exploit} is 2, \textit{Get-Access-Privilege} is 3, and 4 for \textit{Post-Attack}. The suspiciousness of the alert path is the sum of all alert values in the path, as shown in \eqref{pathscore}.

	\begin{equation}
	\label{reliabilityalert}
	Reliability_{a} = Suspicious_{shost} + Suspicious_{dhost} + AlertType
	\end{equation}
	
	\begin{equation}
	\label{pathscore}
	Score_{path} = \sum_{i=1}^n Reliability_{a_i}
	\end{equation}
	
	\section{Evaluation}
	\label{sec:evaluation}
	
	We evaluate our alert correlation algorithm MAAC on real-world datasets. Alerts are needed to test this method, but there is currently no dataset composed entirely of alerts according to \cite{ring2019survey}. Therefore, we first use IDS to generate alerts based on the traffic packet and logs in the dataset. For attacks that cannot be detected by ordinary IDS, for example, we use artificial alerts to assist with testing. We use \textit{python 3} to achieve alert reduction, alert correlation graph generation, graph traversal, and path evaluation. 
	
	\subsection{NLP Evaluation}
	
	MAAC uses Doc2vec to vectorize the alert information field, then integrates similar alerts, and classifies alerts to obtain the category of alerts. We use attack detection rules that can generate alerts as the evaluation set, including snort and suricate rule sets, as well as rules collected from Github. After the classification to four attack stages by the classtype field and manual confirmation, we divide the set into the training set and test set.
	
\begin{figure}[htbp] 
		\centering 
		\includegraphics[width=0.45\textwidth,trim=0 20 0 10,clip]{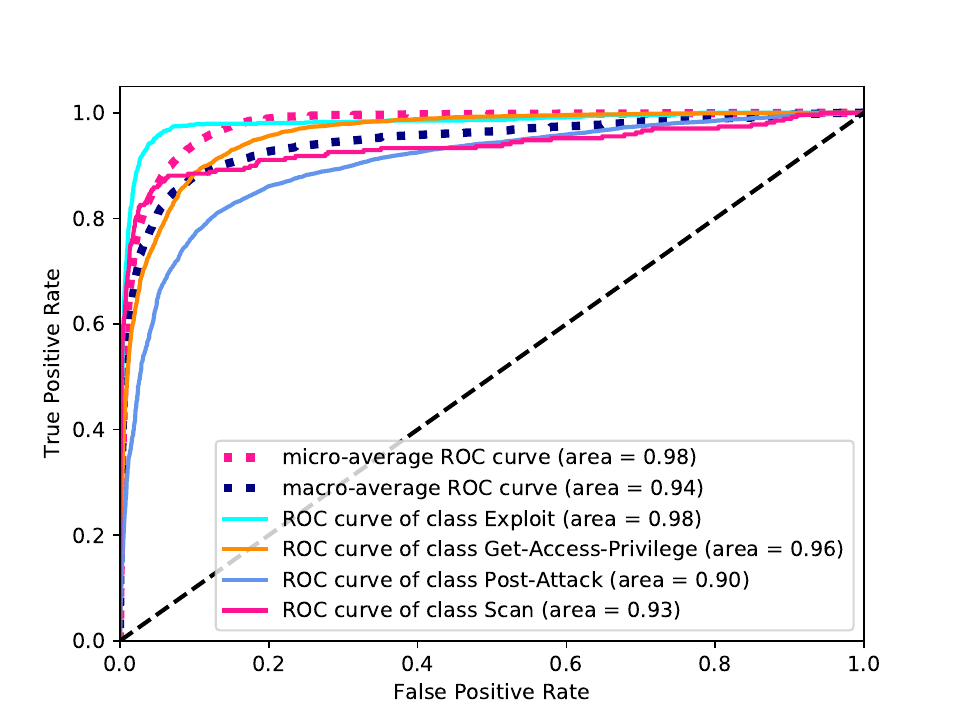} 
		\caption{ROC curve of MAAC's Doc2vec model. The solid lines represent the ROC curves of different categories, and the dotted lines represent the average multi-category ROC curves.} 
		\label{Fig.roc} 
	\end{figure}

	The logistic regression model is used to classify the vectors and obtains the accuracy of 90.56\% and the F1 value of 90.47\%. The receiver operating characteristic (ROC) curve of the vector semantics and classification method is shown in Figure \ref{Fig.roc}. The average Area Under Curve (AUC) value of MAAC using the Doc2vec model and logistic regression algorithm is above 0.9. The classification result shows that MAAC can accurately vectorize and classify alerts. 
	
	\subsection{Real-World Evaluation}
	
	We test MAAC on the DARPA 2000 LLDOS1.0 dataset \cite{darpa2000}, UNB ISCX IDS 2012 dataset \cite{shiravi2012toward}, and NDSec dataset \cite{2017Feature}. 
	
	\textbf{Metrics.} For alert reduction, we use the compression rate (Equation \eqref{compressionrate}) to evaluate the effect of the algorithm. To measure the detection ability and correctness of the path, we use two measurement indicators similar to \cite{ning2002constructing}: ${path\;detect\;rate}$ and ${false\;path\;rate}$. ${Path\;detect\;rate}$ is the the ratio of detected attack paths to all attack paths. ${False\;path\;rate}$ is the rate of reporting wrong paths as attack path. The calculation equations are shown in \eqref{rc}, \eqref{rs}.
	
	\begin{equation}
	\label{compressionrate}
	R_{c} = \frac{alerts\;reduced\;count}{alerts\;total\;count}
	\end{equation}
	
	
	\begin{equation}
	\label{rc}
	{path\;detect\;rate} = \frac{reported\;attack\;path}{all\;attack\;path}
	\end{equation}
	
	\begin{equation}
	\label{rs}
	{false\;path\;rate} =1 -  \frac{correctly\;reported\;path}{all\;report\;path}
	\end{equation}
	
	\textbf{DARPA 2000 LLDOS1.0 dataset.} In this dataset, the attacker first sends ICMP echo-requests to scan the surviving hosts, then attempts the Sadmind Remote-to-Root exploit several times against each host, each time with different parameters. There are 6 exploit attempts on each potential victim host and finally successfully breaks into and controls three hosts (\textit{172.16.115.20}, \textit{172.16.112.50}, \textit{172.16.112.10}) in the network using the Sadmind service vulnerability of the host. Then attacker uploads and installs the Mstream DDoS program to the controlled host, and launches an attack on the target server.

	\textbf{UNB ISCX IDS 2012 dataset.} There are 7 days of network traffic in the dataset, including 4 multi-step attack scenarios. Since all attacks in this dataset can be alerted, we select the network traffic on the third day. This traffic includes attack traffic and normal activity traffic. The attacker first sends a malicious email to all the users in the network with the Meterpreter reverse shell attachment, then \textit{192.168.1.105} (user5) is compromised and communicates with the attacker and scans subnet \textit{192.168.1.0/24} and \textit{192.168.2.0/24} according to the attacker's command. User5 successfully attacks \textit{192.168.2.112} (user12) and connects it back to the attacker's machine. User12 then scans the subnet \textit{192.168.5.0/24}, and finally attacks the target server \textit{192.168.5.123} through SQL injection.

	\textbf{NDSec dataset.} We use the NDSec dataset to show that MAAC still has the ability to mine attack fragments from incomplete alerts. This dataset contains three types of multi-step attacks: Bring-Your-Own-Device, Watering-hole, and Botnet. We use the Botnet scenario to evaluate MAAC. Based on \textit{CVE-2015-2509}, \textit{CVE-2015-5122} and \textit{XSS} vulnerability, the attacker uses \textit{Citadel} botnet malware to infect three hosts. These three bot hosts communicate with the C\&C through HTTP, and the master instructs the bots to execute the attack payload. Bots use SYN to launch DDoS attacks on external destination IP. Two of the bots stole local configuration files and transfer them to an external FTP server.
	
	\subsection{Results in a Nutshell}
	We use snort version 2.9 to generate alerts, with snapshot-2900 ruleset and Emerging-Threats ruleset \cite{EmergingThreats} for network traffic detection and correlate the discovered alerts. It should be noted that MAAC is used to correlate alerts instead of generating alerts, so how to generate attack alerts is beyond the scope of this paper. We use the same ruleset for the three dataset test evaluations. The compression ratios of the three datasets are shown in the table \ref{reductionresult}. The alert compression algorithm of MAAC can reach more than 90\%, which can improve the efficiency of security experts.
	
	\begin{table}[htbp]
		\caption{Reduction Algorithm Result of MAAC in Datasets}
		\centering
		\begin{tabular}{l|l|l|l|l}
			\hline 
			\rowcolor{mygray}
			Dataset         & Packetsize & \begin{tabular}[c]{@{}l@{}}\#Original \\ Alert\end{tabular} & \begin{tabular}[c]{@{}l@{}}\#Reduced \\ Alert\end{tabular}& $R_{c}$      \\ \hline
			DARPA LLDOS 1.0 & 122MB       & 1942         & 164             & 91.56\% \\ \hline
			ISCX-IDS-2012   & 4.2GB       & 22792            & 853             & 96.26\% \\ \hline
			NDSec           & 118MB       & 2259             & 55              & 97.57\% \\ \hline
		\end{tabular}%
		\label{reductionresult}
	\end{table}
	
	\subsection{Case Studies}
	
\begin{figure}[htbp] 
		\centering 
		\includegraphics[width=0.4\textwidth,trim=0 15 0 15,clip]{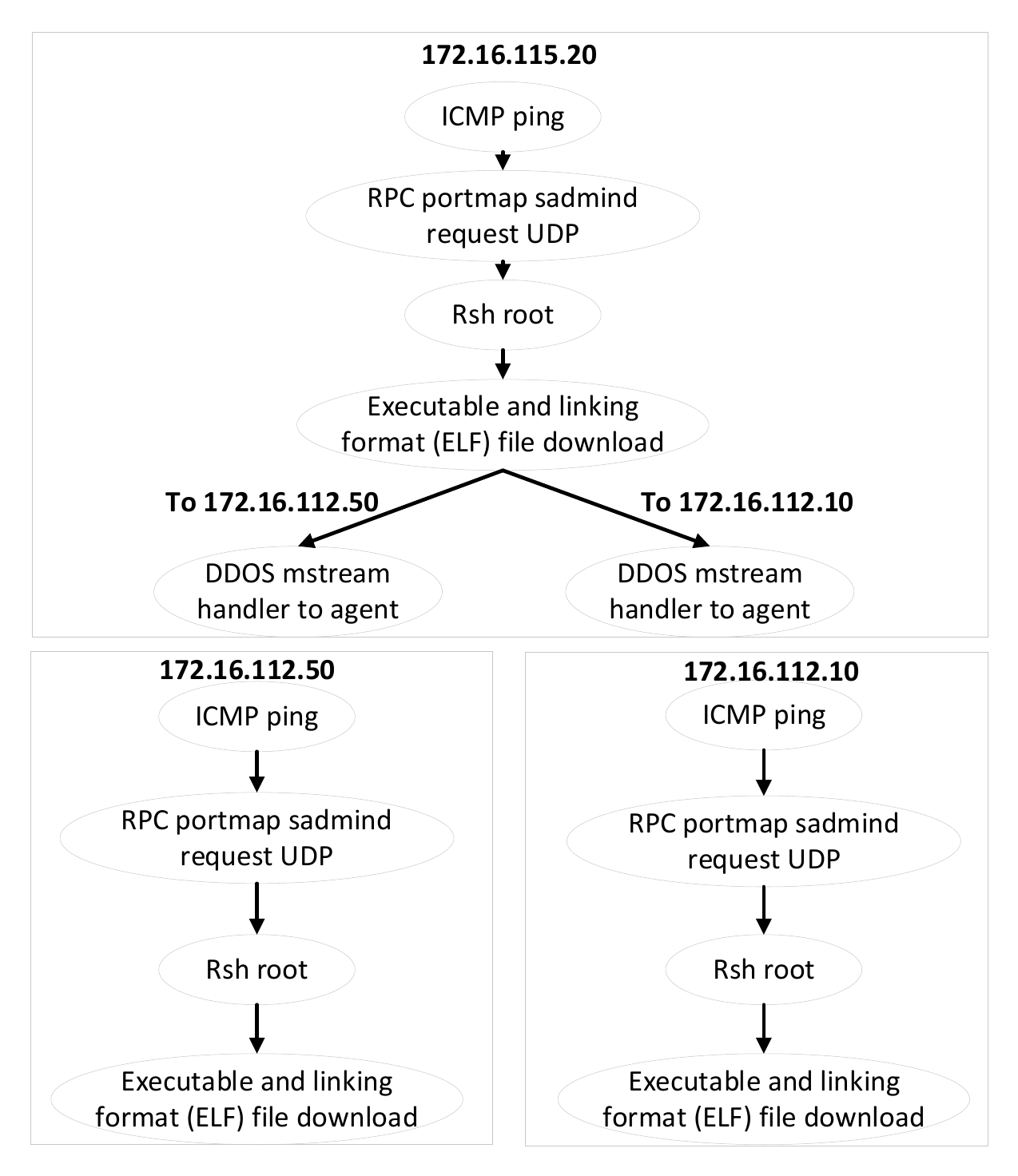} 
		\caption{Multi-step attack paths in DARPA 2000 LLDOS1.0 dataset recovered by MAAC. The attacker (\textit{202.77.162.213}) first executes IPsweep to detect the surviving host, then uses the Sadmind vulnerability to attack live host, gain root privileges, install Mstream DDoS software, and finally issue control commands to the master host (172.16.115.20), which controls the other two hosts (\textit{172.16.112.50}, \textit{172.16.112.10}).} 
		\label{Fig.lldos} 
	\end{figure}

	There are a total of 1942 alerts generated from the DARPA 2000 LLDOS1.0 dataset. After alert reduction, there are 164 alerts, which means that MAAC reduces the number of alerts by 91.56\%. For attack discovery capabilities, MAAC could discover all attack paths. A total of 24 paths are generated, of which were 10 paths with 3 or more nodes. Among the paths, the top 4 paths with higher scores can describe the entire attack process, as shown in Figure \ref{Fig.lldos}. MAAC filtering IP scanning without follow-up attacks and merges 6 Sadmind attacks on victim hosts. MAAC could detect all attack paths, the ${path\;detect\;rate}$ is 100\%. If select the first 4 paths to report as attack paths, the ${false\;path\;rate}$ is 0, if select first 5 paths, the ${false\;path\;rate}$ is 20\%.
	
	
	
	Ning et al. \cite{ning2002constructing} use pre-defined rules to correlate alerts which is regarded as a state-of-art alert correlation approach. We compare MAAC with baseline approach using the DARPA LLDOS1.0 dataset. MAAC automatically clusters, categorizes, and correlates alerts based on the alert semantics. Approach \cite{ning2002constructing} focuses on the verification of real alerts, while MAAC focuses on the reduction of alerts and the ability to find attack paths. Comparison of these two methods is in Table \ref{vsning}. The data indicate that MAAC does not require prior knowledge and can find out the attack path well.
	
	
	\begin{table*}[htbp]
		\caption{Ability to generate alert paths}
		\centering
		\begin{tabular}{c|c|c|c|c|c|c}
			\hline
			\rowcolor{mygray} 
			Dataset                                             & Method    & \#alert  & \#path & \#path detect rate & \#false alert(path) rate & Need Prior Knowledge \\ \hline
			\cellcolor[HTML]{FFFFFF}                          & Ning et al. & 57      & -      & 56.18\%          & 5.26\%                   & yes                  \\ \cline{2-7} 
			\multirow{-2}{*}{LLDOS 1.0 DMZ} & MAAC      & 14      & 3      & 100\%            & 0\%                        & no                   \\ \hline
		\end{tabular}
		\begin{tablenotes} 
			\item {\ \ \ \ \ \ \ \ \ \ \ \ \ \ \ \ \ \ \ \#alert means the alert count in alert graph or path.} 
			\item {\ \ \ \ \ \ \ \ \ \ \ \ \ \ \ \ \ \ \ \#false path rate is 0\% if select top 4 paths as attack paths, 20\% if select top 5 paths.}
		\end{tablenotes} 
		\label{vsning}
	\end{table*}

\begin{figure}[htbp] 
		\centering 
		\includegraphics[width=0.5\textwidth,trim=0 15 0 15,clip]{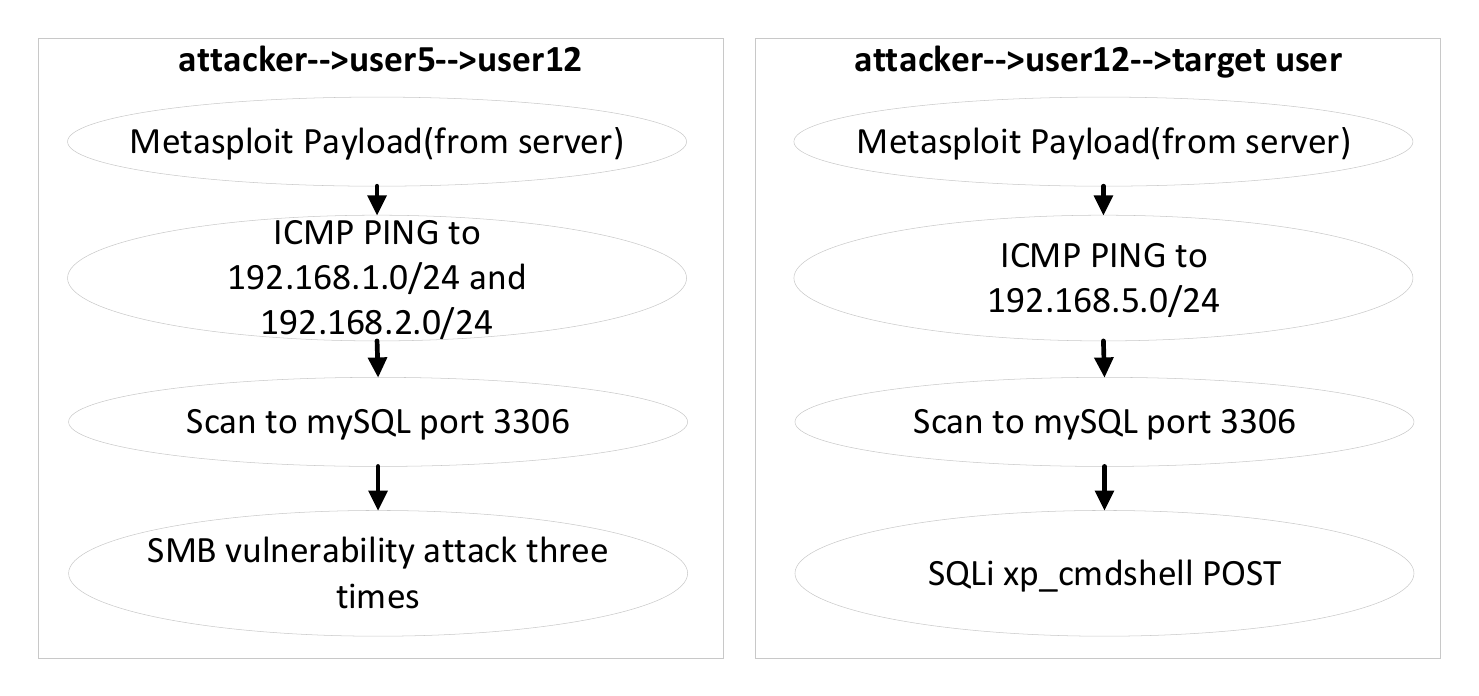} 
		\caption{Multi-step attack paths in UNB ISCX IDS 2012 dataset recovered by MAAC. The first alert in the two paths is the shell from the attacker to 192.168.1.105 and 192.168.2.112, respectively. The remaining alerts are attacks between the internal network hosts.} 
		\label{Fig.iscx} 
	\end{figure}

	After processing the UNB ISCX IDS 2012 dataset, MAAC reduces 22792 alerts to 853 alerts. 43 paths have three or more nodes, 9 of which have more than four nodes. Since user5 and user12 both connect back to the attacker, there are two attack paths, as shown in the Figure \ref{Fig.iscx}.

	In the alerts generated by the NDSec dataset, the XSS attack and final FTP data transfer are not identified, but the communication between three hosts and the C\&C is successfully detected by snort. Under the condition of incomplete alerts, MAAC extracted 20 paths, 12 of which belonged to attack-related paths. The paths include the attacker's exploit methods, the communication between the bot and the C\&C, and the bot's DDoS attack. Besides, the path is identified in which Zeus malware checked the network connection through Google after successfully exploiting.

	\section{Conclusion}
	\label{sec:conclusion}
	
	In this paper, we propose MAAC, a semantics-based method for alert reduction and correlation to solve the problem of the alert explosion. MAAC understands the semantics of alerts and performs alert reduction and correlation. Our method gives possible attack paths and their ranking, as well as the ranking of suspicious hosts. MAAC does not require prior knowledge and can reduce both irrelevant and similar alerts. Through this method, security analysts can reduce their workload and focus on analyzing attack scenarios. We tested the effectiveness of this method in real-world datasets. The results show that MAAC can reduce duplicated alerts by 90\%, and provide critical paths for security analysis. 
	
	%
	
	\bibliographystyle{IEEEtran}  
	\bibliography{ref}
	
\end{document}